\pdfoutput=1

\documentclass[pra,10pt,twocolumn,superscriptaddress, footnoteinbib]{revtex4}
\usepackage{amsmath}
\usepackage{latexsym}
\usepackage{amssymb}
\usepackage{graphics,epstopdf}
\usepackage[colorlinks=true, citecolor=blue, urlcolor=blue ]{hyperref}
\usepackage{epsf,graphics,graphicx}

\newcommand{\ed}{\end{document}}
\newcommand{\beq}{\begin{equation}}
\newcommand{\eeq}{\end{equation}}

\begin{document}
\title{Anomalous thermoelectric properties of a Floquet topological insulator with spin momentum non-orthogonality}
\author{Madhumita Saha}
\email{madhumita.saha91@gmail.com }
\affiliation{Physics and Applied Mathematics Unit, Indian Statistical Institute, Kolkata 700108, India}
	\author{Debashree Chowdhury}
	\email{debashreephys@gmail.com}
	\affiliation{Department of Physics, Ben-Gurion University, Beer Sheva 84105, Israel }

\begin{abstract}
The spin momentum non-orthogonality in 3D topological insulators leads to modification of the spin texture and brings in an out-of-plane spin
	polarization component. Apart from spin texture, the anomalous thermoelectric properties of these materials are worth studying. In this paper, we have pointed out that the off resonant light used to irradiate the surface states, induces a gap, which becomes momentum dependent due to the presence of non-orthogonal terms in the Hamiltonian. Importantly, to maintain the off resonant condition of light, the momentum value should satisfy a bound. Furthermore, the momentum dependent gap causes a topological transition at higher value of momentum, which is important to analyse the unusual double peak structure of the Nernst and electrical conductivities.

\end{abstract}

\maketitle
\section{Introduction}
Topological Insulators(TI)\cite{Hasan,Zhang,Kane} are the unique class of materials with insulating bulk and exotic conducting edge states. The edge states are robust against any non-magnetic impurities, which makes the study of TI non-trivial. In case of a 3D TI, the surface electrons obey a linear dispersion relation, which effectively gives a Dirac cone like structure. However, there exists some anomaly to this usual scenario. In certain TIs, in particular $Bi_{2}Te_{3},$ the Dirac cone deviates from its usual structure \cite{Fu} and becomes warped with increasing energy. For these materials, the usual Fermi surface evolves  like first as a hexagon and then in a snowflake like structure. For such hexagonally distorted Dirac cones, conservation of Berry phase provides an out of plane spin component in the system, which is a crucial ingredient for explaining some unique phenomena such as the enhancement
of interference patterns around crystal defects and exotic
magnetic orders at the surface of the TIs. Furthermore, apart from this warped structures, in \cite{sushmita}, the authors have proposed a higher order warping term in the surface state Hamiltonian to incorporate the non-orthogonality between the in-plane
electron spin and momentum, which cannot be reproduced
by the hexagonal warping term.   

Furthermore, in condensed matter systems, Bloch states and bands corresponds to spatially
periodic Hamiltonians\cite{21}. If the periodicity is extended by adding time periodic perturbations, one gets further informations that Bloch states are unable to carry. Recently, topological phases of periodically driven quantum systems have been characterized by Floquet theory \cite{Tahir,debabrata,Torres,Torres 1,Torres 2,Torres 3}. We restrict ourself to the off resonant condition\cite{Tahir,Tahir1}, which can be obtained when the frequency of the optical field is much larger than the bandwidth of the system. The off resonant light does not signify any electron transitions between different energy bands but the band structure of the system gets renormalized via a processes of the virtual photon absorption and emission. The floquet theory is analysed in different systems like photonic crystals \cite{22},graphene\cite{29}, silicene\cite{Ezawa} and TIs\cite{23,24}. In this paper, we have discussed the floquet theory of TI with spin momentum unlocking terms, where any electronic transitions in the system are ignored. This modifies the band structure and introduce a gap by breaking time reversal symmetry\cite{Cayssol}. We have shown that when the non-orthogonal terms are present, surprisingly the gap opened as a consequence of light, becomes momentum dependent. The momentum dependence of the gap is crucial for our case as it induces a bound on the values of the momentum(to satisfy the off resonant condition). On the other hand, the Berry curvature mediated thermoelectric transport\cite{Tahir,takehito} is a topic of great recent interest. Here a temperature gradient can mediate thermoelectric effects \cite{Uchida}.  Thermoelectric coefficients can be computed using its relation with the Berry phase and curvature. It was first shown for the case of 2D electron gas \cite{Xiao}. The results are then verified in a number of experiments. In \cite{takehito}, the authors have analysed the Berry-curvature mediated heat transport on the surface states of a 3D TI attached to a ferromagnet. This leads to the interesting issue of
controlling the thermoelectric effects \cite{16,17,18,19} via irradiating the surface of TIs with circularly polarized off-resonant light \cite{20}. As far as our knowledge goes, the effect of the spin momentum non-orthogonality on the light induced gap is missing in the literature. Our goal in this paper is to explore the effects of the spin momentum non-orthogonal terms induced momentum dependent gap on the Nernst and electrical conductivities of the system, which are experimentally
accessible at different temperatures due to the
tunable band gap \cite{ex}.

The organization of the paper is as follows: In section II, we have indicated our model Hamiltonian and how the off resonant light produces an effective Hamiltonian is discussed. Sec III, contains the evaluation of the thermoelectric transport coefficients . In sec. IV we have analysed the numerical plots for different thermoelectric coefficients. Finally, we conclude in sec V.  

\section{The system and Floquet theory}
We can start with the surface state Hamiltonian of a 3D topological insulator with higher order warping  terms as follows \cite{debabrata,sushmita,Fu,Re,Book,Ha}
\begin{eqnarray}\label{1}
H(k) &=& \hbar v(k_x\sigma_y- k_y\sigma_x)
+{\lambda\over{2}}(k_{+}^3+k_{-}^3)\sigma_z \nonumber\\&+& i\zeta (k_{+}^5\sigma_+ - k_{-}^5\sigma_-),
\label{hamil}
\end{eqnarray}
where $k_\pm=k_x \pm ik_y$, $\sigma_\pm=\sigma_x \pm i\sigma_y$ and the $\sigma_i$ are Pauli matrices.  The form of $H(k)$ is suitable for describing the [111] surface band structure near the $\Gamma$ point in the surface Brillouin zone of  $\rm Bi_2Te_3$ family TIs and
it is invariant under time-reversal and $C_{3v}$ symmetries. In (\ref{1}), the first term is the usual spin momentum locking term, which is the well known Bychkov–Rashba Hamiltonian for 2D electrons
with spin orbit splitting \cite{Rashba}. The second term is known as the hexagonal warping term with $\lambda$ as the hexagonal warping parameter. We should note here that although Fu \cite{Fu} in his observation have a $k^{2}$ dependence of the parameter, but for calculational simplicity, we consider $\lambda$ as constant. The last term is the higher order warping term, with $\zeta$ as the coupling parameter\cite{sushmita}.

Let us now irradiate the surface of the 3D TI with a beam of polarized optical field as $E(t)=E_{0}(\cos(\omega t),-\sin(\omega t))$, where $E_{0}$ and $\omega$ are the amplitude and frequency of the optical field. The electric field is associated with the vector potential as $\vec{A}(t)=A_{0}(\sin(\omega t),\cos(\omega t))$ with $A_{0}=-E_{0}/ \omega$.  As it is obvious that the light matter interaction can be incorporated in the Hamiltonian via the Peierls substitution, which replaces the momentum $\hbar k_{i} \rightarrow \hbar k_{i}+eA_{i},$ where $\vec{A}(t+T)=\vec{A}(t)$ with $T=2\pi/ \omega$ as the periodicity. Considering the time scale used in the system as large compared to $T$, it is possible to apply the elegant mechanism, known as the Floquet theory \cite{debabrata,1,2,41}.

In the present analysis the time dependent gauge field can be written as $A(t)$=$A_0 (\sin (\omega t),\cos (\omega t)),$ which effectively modifies the $k_x$ and $k_y$ as
$k_x$=$k_x+ \frac{e A_0 \sin(\omega t)}{\hbar}$;$k_y$=$k_y+ \frac{e A_0 \cos(\omega t)}{\hbar}.$

Substituting this modified $k_x$ and $k_y$ in (\ref{1}), the new Hamiltonian becomes,
\begin{eqnarray}
\mathcal{H}(k,t) &=& \hbar v(k_x\sigma_y- k_y\sigma_x)
+{\lambda\over{2}}(k_{+}^3+k_{-}^3)\sigma_z\nonumber\\&& + i\zeta (k_{+}^5\sigma_+ - k_{-}^5\sigma_-)+ \mathcal{V}(t)+\mathcal{O}(A^2, A^3)\nonumber\\&=&\mathcal{H}_{0}+ \mathcal{V}(t),
\label{hamil1}
\end{eqnarray}

where, $\mathcal{V}(t)$=$e v (A_x\sigma_y-A_y\sigma_x)+ \frac{3 e \lambda}{2 \hbar}[k_{+}^2(A_x+i A_y)+ k_{-}^2(A_x-i A_y)]\sigma_z+ \frac{5 i\zeta e}{\hbar}[k_{+}^4(A_x+i A_y)\sigma_{+}-k_{-}^4(A_x-i A_y)\sigma_{-}] $

The role of the off resonant light can only be noticed in the static effective Hamiltonian in terms of the evolution operator $U$ \cite{41}
\begin{eqnarray}
\mathcal{H}_{eff}(k)=\frac{i\hbar}{T}\log U ,
\end{eqnarray}
where,
\begin{align}
U =T_{t} \exp[\frac{1}{i\hbar}\int^{T}_{0}\mathcal{H}(k,t)dt]
\end{align}
with $T_{t}$ as the time-ordering operator. The effective Hamiltonian $\mathcal{H}_{eff}$ basically describes the dynamics of the system on the time scale much longer than a period $T$, thus the response can be well described by an average over a period of oscillation. Use of Floquet theorem produces the matrix elements of the time-dependent Floquet Hamiltonian \cite{zhou,zhouprb15,38,39,40,41}
\begin{eqnarray}
\mathcal{H}^{m,m^{'}}_{F}=\mathcal{H}_{0}\delta_{m,m^{'}}+m\hbar \omega \delta_{m,m^{'}}+\mathcal{H}^{'}_{m,m^{'}}
\end{eqnarray}
where $\mathcal{H}^{'}_{m,m^{'}}=V_{n}=\frac{1}{T}\int^{T}_{0}\mathcal{V}(t)e^{i(m-m^{'})t}dt=\frac{1}{T}\int^{T}_{0}\mathcal{V}(t)e^{in\omega t}dt.$ 
Considering  the laser induced field purturbatively, one can have the general quasistatic Floquet Hamiltonian as,
\begin{eqnarray}
\mathcal{H}_{F}\simeq \mathcal{H}_{0}+\sum^{\infty}_{n=1}(-1)^{n-1}\frac{1}{n\hbar \omega}\frac{[V^{n}_{-1},V^{n}_{1}]}{[(n-1)!\hbar \omega]^{2(n-1)}}.
\end{eqnarray}

Using these elements one can write the static time independent effective Hamiltonian upto order $1/\omega$ as,
\begin{eqnarray}
\mathcal{H}_{eff}=\mathcal{H}_{0}+\frac{[V_{-1},V_{+1}]}{\hbar \omega}.
\label{eff-hamiltonian}
\end{eqnarray}
In the present case we can calculate the corresponding elements as follows,
\begin{eqnarray}
V_{-1}&=& i\alpha (i\sigma_x+\sigma_y)+i\beta k_{+}^2\sigma_z+\gamma k_{+}^4(\sigma_x+i \sigma_y)\nonumber\\
V_{1}&=& i\alpha (i\sigma_x-\sigma_y)-i\beta k_{-}^2\sigma_z+\gamma k_{-}^4(\sigma_x-i \sigma_y)
\end{eqnarray}
Where $\alpha=e v A_0/2$; $\beta=\frac{3 e \lambda A_0}{2\hbar}$; $\gamma=\frac{-5\zeta e A_0}{\hbar}.$
Here we have considered terms up to first order of the light potentials. From eqn. (\ref{eff-hamiltonian}),  the effective static Hamiltonian can be obtained as
\begin{eqnarray}
\mathcal{H}_{eff}&=&\hbar v[(k_{x}+\mathcal{K}_{1}(k_{x},k_{y}))\sigma_{y}-(k_{y}+\mathcal{K}_{2}(k_{x},k_{y}))\sigma_{x}]\nonumber\\&&+\frac{\lambda}{2}(k^3_{+}+k^{3}_{-})\sigma_{z}+\Delta_{\omega} \sigma_{z}
\label{Eff-Hamil},
\end{eqnarray}
where \begin{align}
\mathcal{K}_{2}(k_{x},k_{y})&=-\frac{i\zeta}{\hbar v}(k_{+}^5-k_{-}^5)+2 a
k_x k_y(\alpha-\gamma (k_{x}^2+k_{y}^2)^2),\nonumber\\
\mathcal{K}_{1}(k_{x},k_{y})&=-\frac{\zeta}{\hbar v}(k_{+}^5+k_{-}^5)- a
(k_{x}^2- k_{y}^2)(\alpha-\gamma (k_{x}^2+k_{y}^2)^2),
\end{align}  with  $a=\frac{4\beta}{\hbar^2 \omega v}.$
Importantly, we have \beq \Delta_{\omega} = \frac{4(\alpha^{2}-\gamma^{2}(k_{x}^2+k_{y}^2)^4)}{\hbar \omega} ,\eeq which effectively provides the gap between the conduction and valance band. It is to be noted here that the gap $\Delta_{\omega}$ depends on both the light amplitude and the momentum $k=\sqrt{k_{x}^{2}+k_{y}^{2}}.$ Although the gap opening is not a new thing in the Floquet theory, the $k$ dependence of the gap makes the present analysis non-trivial. This is due to the presence of a $\zeta$ dependent term in the Hamiltonian. Importantly, this is one of the main observations of our paper. The $k$ dependence of the gap imposes some restrictions on the values of the momentum, which are allowed by the off resonant condition. 

Here we consider the off resonant condition\cite{Tahir,debabrata}, which can be obtained when the frequency of the optical field is much larger than the bandwidth of the system. The off resonant light does not signify any electron transitions between different energy bands but the band structure of the system gets renormalized via a processes of the virtual photon absorption and emission.The off-resonant condition thus demands that the frequency of the light should be much larger than any scale of the system. FIG 1 shows the exact nature of the light induced gap with $k.$ The gap decreases with momentum and becomes zero for a particular value of $k$ and then changes its sign. This is the characteristic feature of the topological transition. The nature of the variation of the gap with momentum plays the key role for peculiar nature of the thermoelectric transport coefficients, which we will discuss in the next section. 
 
\begin{figure}
	{\centering \resizebox*{8cm}{9cm}{\includegraphics{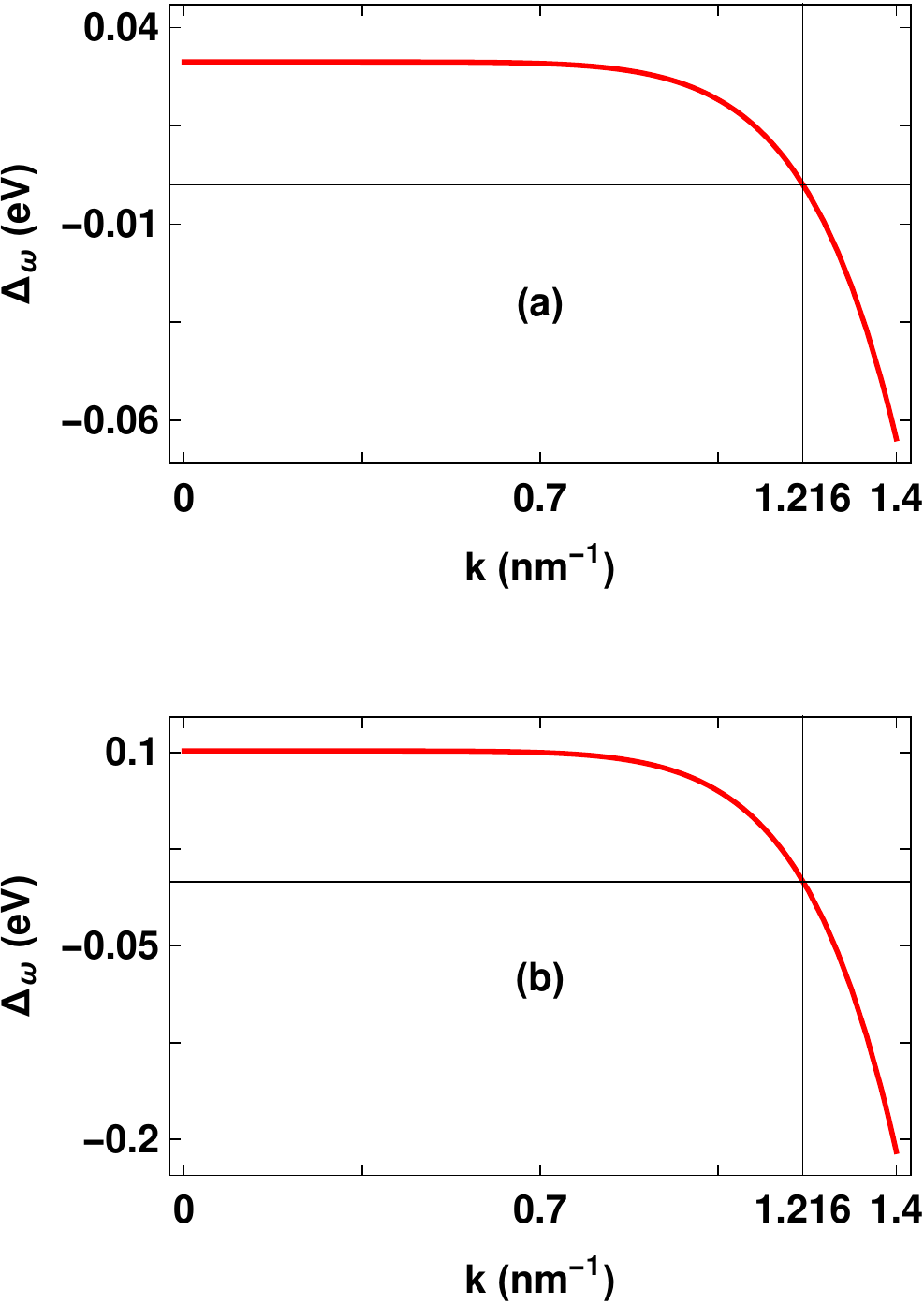}}\par}
	\caption{(Color online). Light frequency dependent band gap $\Delta_{\omega}(eV)$ is plotted with $k(nm^{-1})=\sqrt{k_x^{2}+k_y^{2}}$. In fig. (a) light intensity $e v A_0$=0.5 and (b) $e v A_0$=0.9. With increasing light strength, band gap
		increases. After a critical value of $k$, band gap crosses the zero line and it
		becomes negative for both the cases. We have considered off-resonant condition
		that leads to a bound in $k$ value.}
	\label{f1}
\end{figure}
To obtain the eigen energies of Hamiltonian (9), we can write 
$k_{+}^3+k_{-}^3=2(k_{x}^3-3k_x k_{y}^2)=2 k^3 \cos(3\theta)$ and 
$\mathcal{H}_{eff}$ can be written in matrix form as
\begin{widetext}	
\[H_{eff} = 
\begin{bmatrix}
\lambda k^3\cos[3\theta]+\Delta_{\omega}       & 
\hbar v (-i k_{-}+\frac{2 i\zeta}{\hbar v}k_{+}^5+i a k_{+}^2(\alpha-\gamma (k_{x}^2+k_{y}^2)^2) \\
\hbar v (i k_{+}+\frac{-2 i\zeta}{\hbar v}k_{-}^5-i a k_{-}^2(\alpha-\gamma (k_{x}^2+k_{y}^2)^2)      
& -(\lambda k^3\cos[3\theta]+\Delta_{\omega}).  \\

\end{bmatrix}
\]
\end{widetext}
Denoting
$\Delta(k,\theta)=\lambda k^3 \cos[3\theta]+ \Delta_{\omega}$ and
$A(k,\theta)=\hbar^2 v^2[k^2-\frac{4 \zeta k^6}{\hbar v}(2 \cos^2[3\theta]-1)-
2 a k^3 \cos[3 \theta](\alpha-\gamma k^4)+(2\zeta/(\hbar v))^2 k^{10}+(4\zeta a/(\hbar v))k^7 \cos[3\theta] (\alpha-\gamma k^4)+a^2 k^4 (\alpha-\gamma k^4)^2]$,
 the final expression of energy values can be written as
\beq \epsilon_{s}(k)=s \sqrt{\Delta(k, \theta)^2+A(k, \theta)}.\eeq
The eigen-states of the Hamiltonian (6) are
\[
\chi_s(k)= C_s
\begin{bmatrix}
1        
\\
\frac{\hbar v (i k_{+}+\frac{-2 i\zeta}{\hbar v}k_{-}^5-i a k_{-}^2(\alpha-\gamma (k_{x}^2+k_{y}^2)^2)}{\epsilon_s+\Delta(k,\theta)}
\end{bmatrix}
\]

where $C_{s}^2=[1+\frac{A(k,\theta)}{(\epsilon_s+\Delta(k,\theta))^2}]^{-1},$ with $s=\pm 1.$
\begin{figure}
        {\centering \resizebox*{11.5cm}{7cm}{\includegraphics{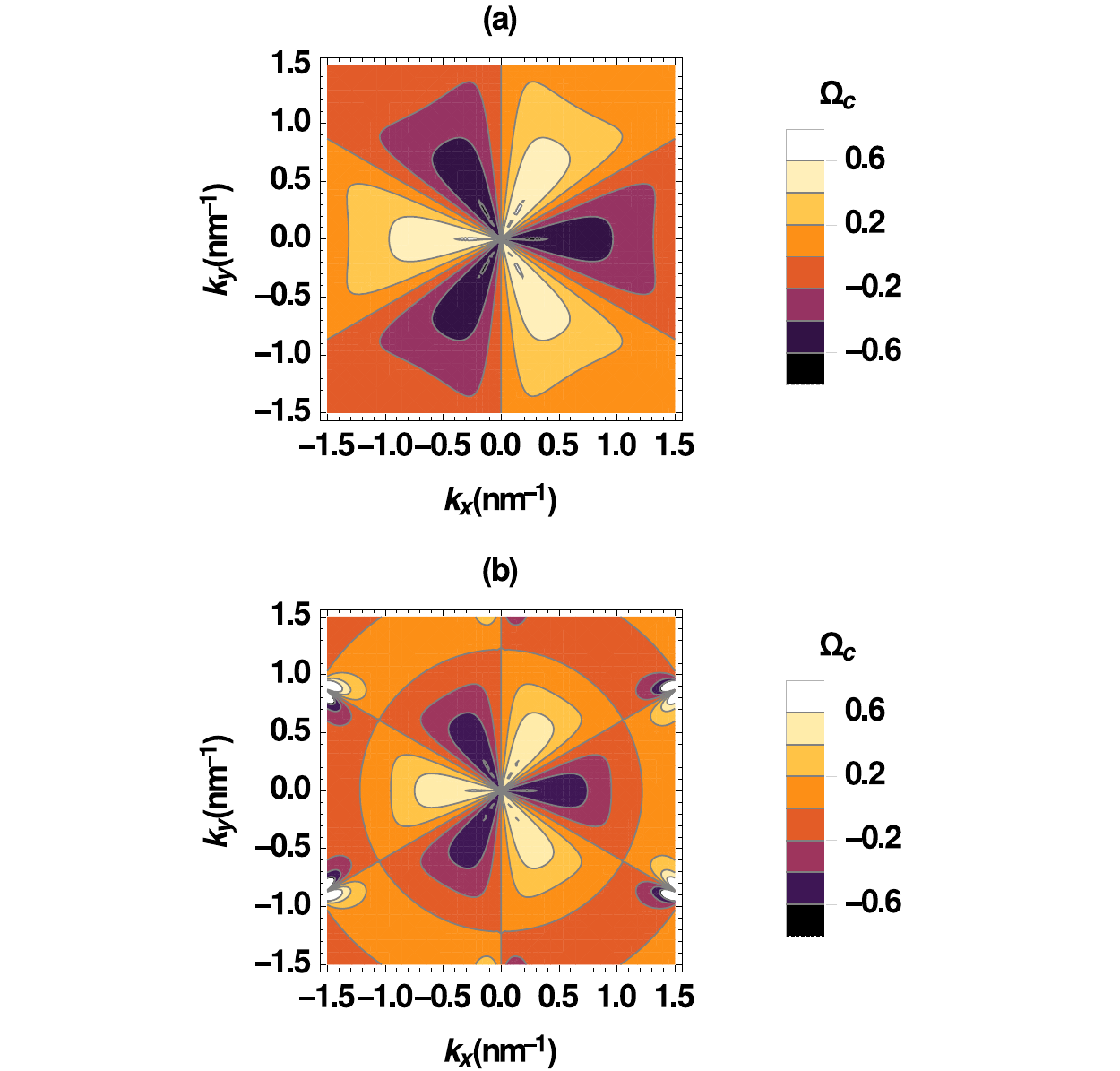}}}
        \caption{(Color online).Berry Curvature($\Omega_{c}$) as a function of momentum $k_x(nm^{-1})$ , $k_y(nm^{-1})$ in the surface states Brillouin zone  .In (a)  the warping $\lambda$ = 0.2 $eV nm^3$ and in (b) $\lambda$ = 0.2 $eV nm^3$ and $\zeta$=-0.015 $eV$ $nm^5$ in absence of optical field }
        \label{f0}
\end{figure}

\begin{figure}
        {\centering \resizebox*{11.5cm}{7cm}{\includegraphics{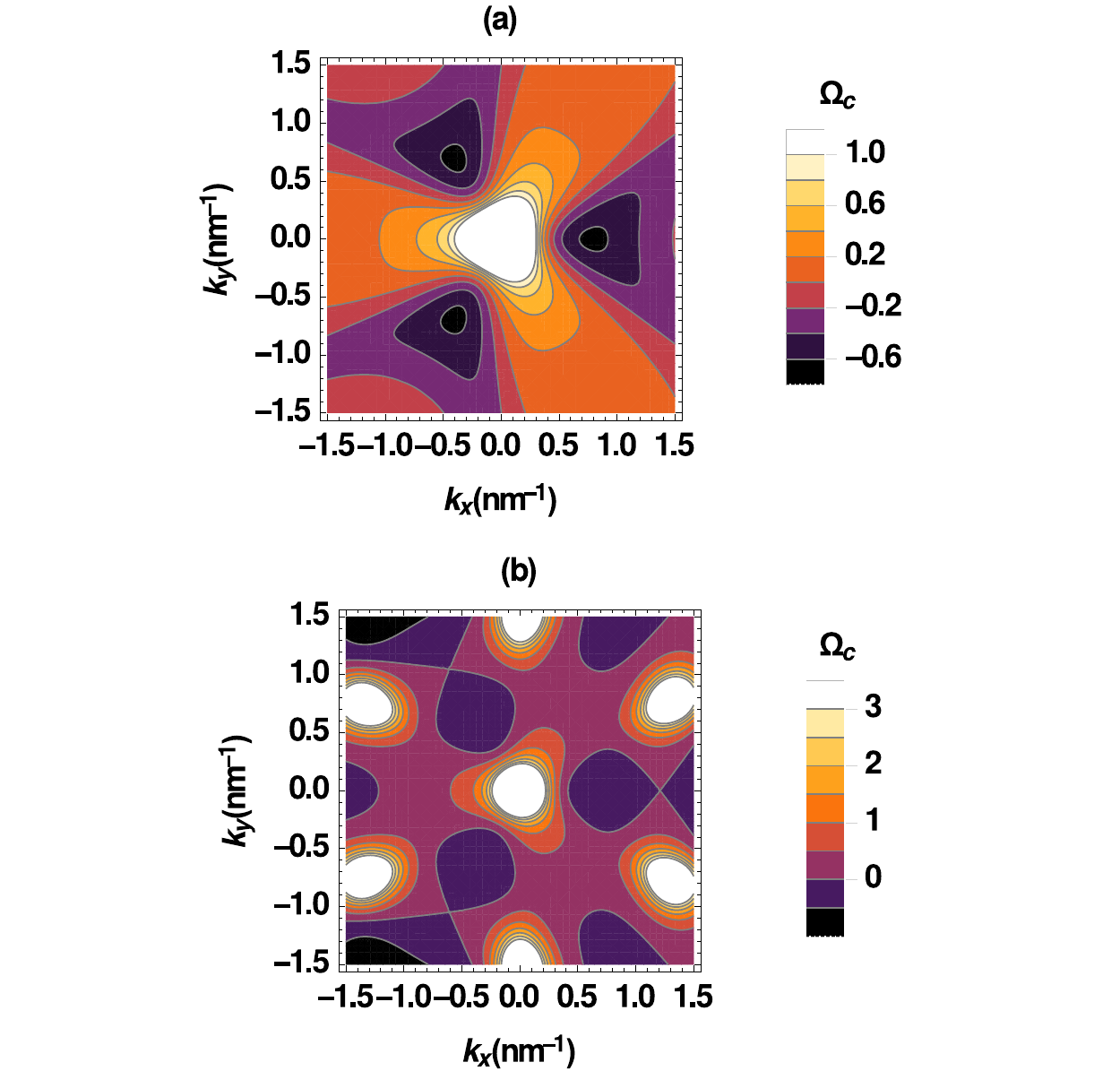}}}
        \caption{(Color online).Berry Curvature($\Omega_{c}$) as a function of momentum $k_x(nm^{-1})$ , $k_y(nm^{-1})$ in the surface states Brillouin zone. In (a)  the warping $\lambda$ = 0.2 $eV nm^3$ and in (b) $\lambda$ = 0.2 $eV nm^3$ and $\zeta$=-0.015 $eV$ $nm^5$ in presence of optical field ($e v A_0$=$0.5$)}
        \label{f00}
\end{figure}
Through out the paper we have considered $\hbar\omega$=$8$ $eV$, fermi velocity
$v$=$0.5\times 10^6$ meter/sec and lattice spacings=0.4 nm. We choose,
$e v A_0$=0.3 eV, 0.5 eV and 0.9 eV respectively and $\Delta_{\omega}<|0.4|$ to satisfy the off resonant condition. Using these values in eqn. (11), we can have the exact value of the bound in momentum. 
\section{Thermoelectric coefficients}
\begin{figure}
        {\centering \resizebox*{9.5cm}{6cm}{\includegraphics{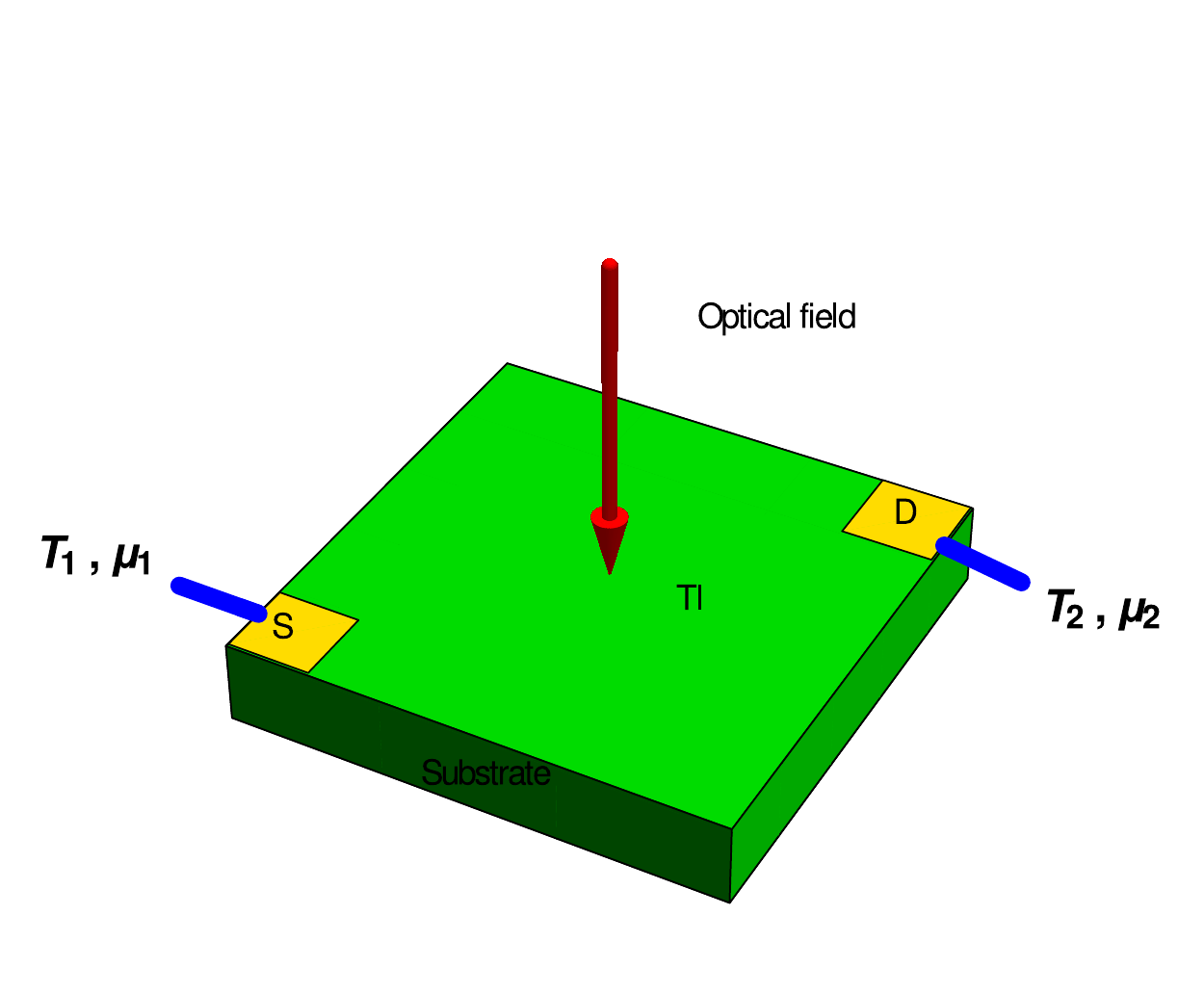}}}
        \caption{(Color online)Schematic diagram of a irradiated TI surface state in presence of temperature gradient(${\nabla T}=(T_{1}-T_{2})$) and voltage bias${\nabla \mu}=(\mu_{1}-\mu_{2})$.}
        \label{setup}
\end{figure}
The study of the thermoelectric coefficients for the surface states of the topological insulators is a topic of great recent interest. We discuss here the scenario of transverse heat transport on the surface of the
3D floquet topological insulator. The heat current in the presence of a weak electric field $E$ and a small thermal
gradient $\nabla T$ can be written as \cite{takehito,Tahir,Tahir1}
\beq j_{Q}= T{\bf \alpha}\cdot {\bf E}-{\bf \kappa} \cdot {\bf \nabla T},\label{jq}\eeq where $\alpha$ and $\kappa$ are the Nernst and heat conductivity tensors respectively.
These quantities can be calculated as \cite{takehito,Xiao}
$\alpha _{xy}  = \frac{{k_b e}}{h}c_1 ,\kappa _{xy}  =  - \frac{{k_b^2 T}}{h}c_2 $ where
\begin{eqnarray}
c_i  = \int_{}^{} {\frac{{d{\bm{k}}}}{{(2\pi )^2 }}} \sum\limits_\tau  {\Omega _\tau  } \int_{\varepsilon_{k \tau} - \mu }^\infty  {d\varepsilon (\beta \varepsilon )^i \frac{{\partial f(\varepsilon )}}{{\partial \varepsilon }}},
\end{eqnarray}
where $\tau $ is the band index and $\Omega_{\tau}$ and $f(\varepsilon )$ are the Berry curvature and Fermi function respectively.
It is to be mentioned here that the Berry curvature plays a significant role in the computation of these coefficients. The Berry curvature renormalises directly the intrinsic Hall conductivity, which can be represented as an integral involving the Berry curvature over the 2D Brillouin zone. Apart from the Hall conductivity one needs the knowledge of the Nernst conductivity as well. These two differ in a sense that, the Nernst one is dependent of both the Berry curvature and the entropy generation around
the Fermi surface. As a consequence, Nernst conductivity is much more sensitive to changes of the Fermi energy and temperature. The Nernst conductivity can be obtained as \cite{takehito,Xiao,Tahir,Tahir1}
\begin{eqnarray}
 \alpha_{x y}&=&- \frac{k_b e}{h} \int \frac{d{\bf k}}{(2\pi)^2}\sum_{\tau}\Omega_{\tau}[\beta(\epsilon_{k\tau}-\mu)f(\epsilon_{k\tau}-\mu)\nonumber\\&+&ln(1+e^{-\beta(\epsilon_{k\tau}-\mu)})],\label{8}
 \end{eqnarray}
 with $f$ is the Fermi function and $\mu$ is the chemical potential. It is to be mentioned here that the  electric field presented in eqn. (\ref{jq}) can be represented as $E={\bf \nabla\mu} =(\mu_{1}-\mu_{2}),$ where $\mu_{1}$ and $\mu_{2}$ are the two chemical potentials of the left and right reservoirs as shown in FIG. 4.
In a similar fashion one can define the thermal conductivity tensor component as \cite{takehito,Xiao,Tahir,Tahir1}

\begin{eqnarray} 
\kappa_{x y} &=& \frac{k_{b}^2 T}{h} \int \frac{d{\bf k}}{(2\pi)^2}\sum_{\tau}\Omega_{\tau}[\frac{\pi^2}{3}+\beta^2(\epsilon_{k\tau}-\mu)^2 f(\epsilon_{k\tau}-\mu)\nonumber\\&-&[ln(1+e^{-\beta(\epsilon_{k\tau}-\mu)})]^2-2{\bf Li}_2[1-f(\epsilon_{k\tau}-\mu)]]\label{9},
\end{eqnarray}
where $Li_{2}(x)$ is the polylogarithm function. Similarly, the electrical or Heat conductivity can also be represented in terms of the Berry curvature as \cite{takehito,Xiao,Tahir,Tahir1},
\beq \sigma_{x y}=\frac{e^2}{2 \pi h}\int \frac{d{\bf k}}{(2\pi)^2}\sum_{\tau}\Omega_{\tau}f(\epsilon_{k\tau}-\mu) \label{10}\eeq
 One important thing to be mentioned here is that here the integration limits are not the whole first BZ, rather it is restricted by the off resonant condition as discussed before. The eqns in (\ref{8}), (\ref{9}) and (\ref{10}), show the involvement of Berry curvature. Thus it is important to calculate the Berry curvature for the system at hand, for which we can write the effective Hamiltonian in eqn. (9), in the following form
\beq\mathcal{H}_{eff}=\Lambda.\sigma,\eeq with 
 $\Lambda_{x}$=$\hbar v [-k_y-\frac{2\zeta}{\hbar v}(k_{y}^5-10 k_{x}^2k_{y}^3+
 5 k_{x}^4 k_y)-2ak_xk_y(\alpha-\gamma(k_{x}^2+k_{y}^2)^2]$; $\Lambda_{y}$=$\hbar v [k_x-\frac{2\zeta}{\hbar v}(k_{x}^5-10 k_{x}^3k_{y}^2+5 k_{x} k_{y}^4)-a(k_{x}^2-k_{y}^2)(\alpha-\gamma(k_{x}^2+k_{y}^2)^2]$ and 
 $\Lambda_{z}$=$\lambda (k_{x}^3-3k_xk_{y}^3)+\frac{4\alpha^{2}}{\hbar \omega}-
 \frac{4\gamma^{2}}{\hbar\omega}(k_{x}^2+k_{y}^2)^4.$ Let us introduce a unit vector $\hat{z}$ along the $z$ direction. Using the definition of the Berry curvature as,
 \begin{eqnarray}
\vec{\Omega}_{c}(\vec{k})=\hat{z}\frac{\vec{\Lambda}.(\partial_{k_{x}}\vec{\Lambda}\times \partial_{k_{y}}\vec{\Lambda})}{2\epsilon^{3}(k_{x},k_{y})}
\end{eqnarray}
one obtains the final form of Berry curvature  for the conduction band in absence of optical field as follows, \cite{zhou},
\begin{widetext} 
 \beq \Omega_c(k)=-\frac{2\lambda(k^3 \cos(3\theta)(\hbar^2 v^2+12\hbar v\zeta k^4+20\zeta^{2}k^8) }{2(\hbar^2
 	v^2(k^2-\frac{4\zeta k^6}{\hbar v}(2\cos^{2}(3 \theta)-1)+(2\zeta/(\hbar v))^2 k^{10})+\lambda^2 k^6 \cos^2(3\theta))^{\frac{3}{2}}}\label{Ome}.\eeq
\end{widetext}
Importantly, the dependence of the Berry curvature on the $\lambda$ and $\zeta$ is evident from the expression in (\ref{Ome}). Berry curvature for the valence band and conduction band are related as $\Omega_{v}$=$-\Omega_{c}$. It is to note here that the Berry curvature is symmetric in the absence of light and any higher order warping terms, which is obvious from the FIG 2. Once the light is turned on, the scenario is completely changed and we obtain some gap opening(see FIG 3). FIG 3, shows the exact pattern of the Berry curvature when we turn on only $\lambda$ term and both $\lambda$ and $\zeta$ terms respectively.  We can comment here that when both the coupling parameters are present in the system we obtain a maximum values of the Berry curvature in two different positions, unlike the case with only $\lambda$ term, where the maximum value of the Berry curvature occurs at $k_{x}=k_{y} = 0. $ But the presence of the $\zeta$ term generates another peak in the Berry curvature. This is related to the variation of the light induced gap with momentum, which effectively influences the thermoelectric transport coefficients enormously. 

\begin{figure}
	{\centering \resizebox*{8cm}{9cm}{\includegraphics{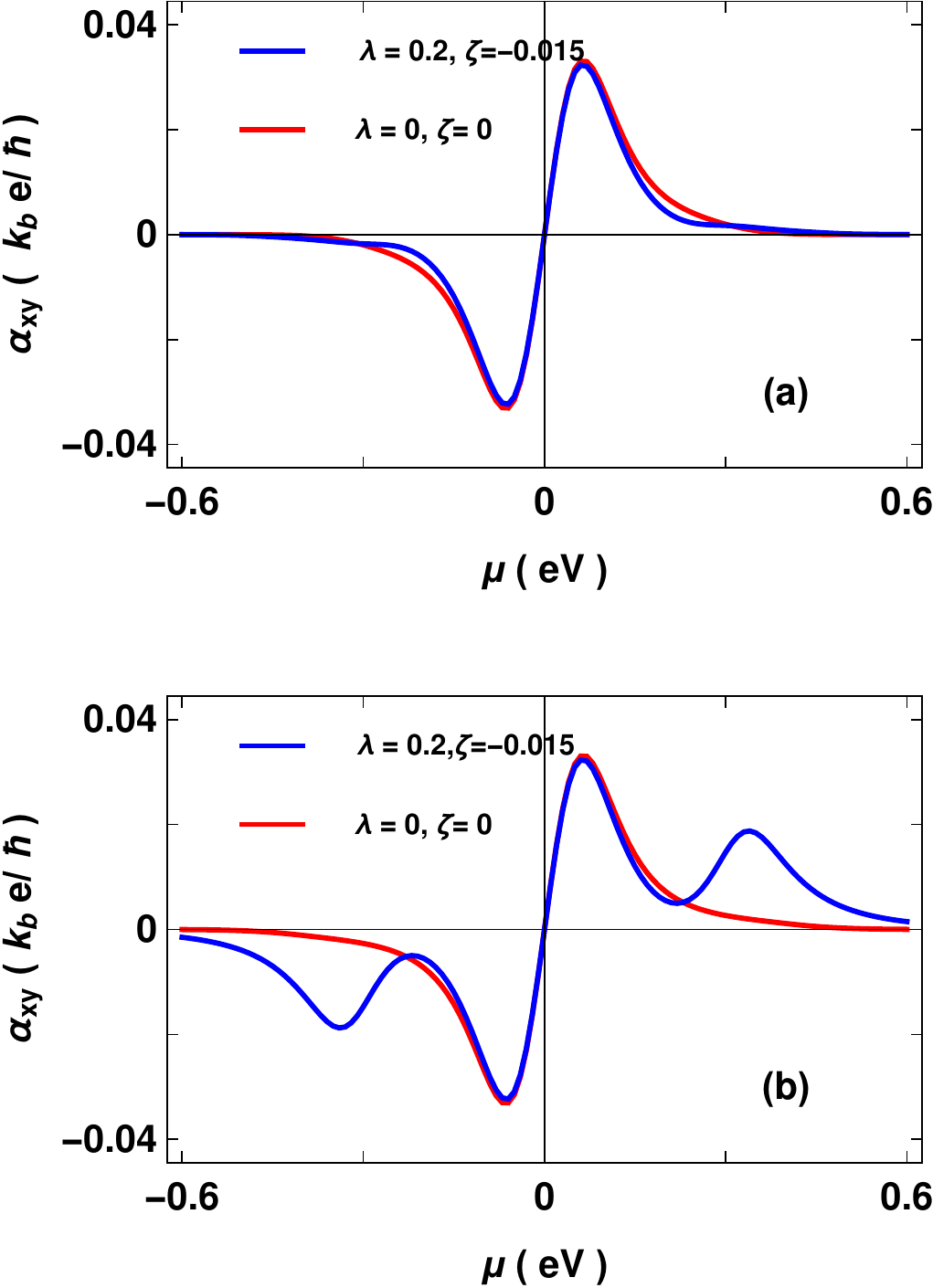}}\par}
	\caption{(Color online). Nernst conductivity $\alpha_{xy}$ in unit of ($k_b e/\hbar$) is plotted with $\mu(eV )$ in absence and presence of higher order
		spin-orbit coupling term. Here the optical field $e v A_0$ is fixed at 0.5 eV
		and the temperature $T$=300K. In fig (a) when $k$ does not cross the zero line
		(b) when $k$ crosses the zero line and band gap is negative.}
	\label{f2}
\end{figure}

\begin{figure}
	{\centering \resizebox*{8cm}{9cm}{\includegraphics{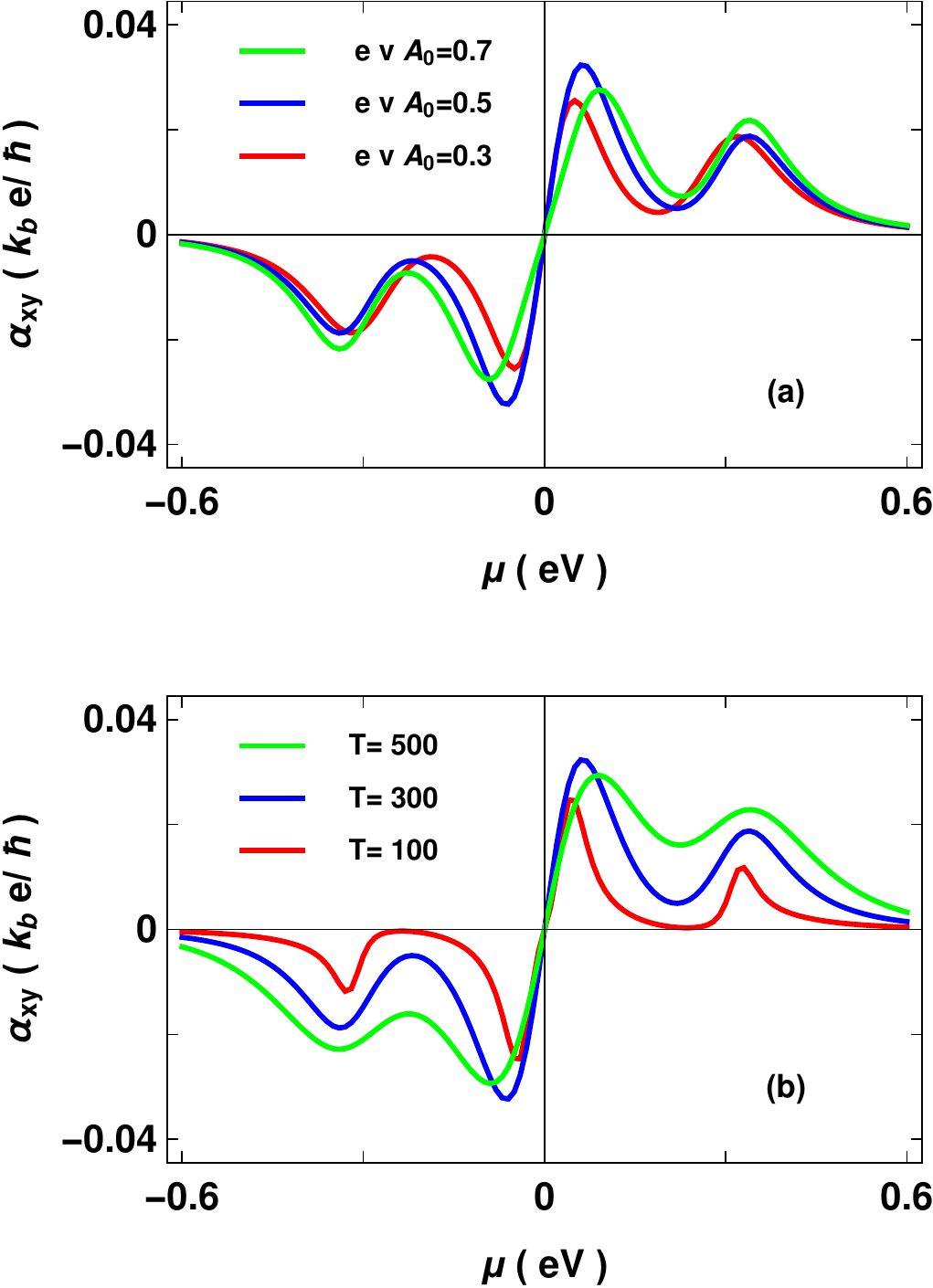}}\par}
	\caption{(Color online). Nernst conductivity $\alpha_{xy}$ in unit of $(k_b e/\hbar)$ vs $\mu$ in unit of electron-volt in presence of higher order warping term$\lambda$=0.2 eV $nm^3$ and $\zeta$=-0.015 eV $nm^5$. In fig.(a), the plot is
		for different temperature and for a fixed optical field $e v A_0$=0.5. In fig.(b), the plot is for different optical field with a fixed temperature $T$=300K.    }
	\label{f3}
\end{figure}

Returning back to our discussion on the thermoelectricity, it is now important to discuss the relations between different coefficients. 
 Thermal conductivity coefficient($\kappa_{i j}$) and electrical conductivity coefficient
 $\sigma_{i j}$ are related in the following way\cite{Tahir,Tahir1},
 \beq \kappa_{i j}=\frac{\pi^2 k_{b}^2 T}{3 e^2}\sigma_{i j} \eeq
 Nernst conductivity \cite{Tahir,Tahir1},
 \beq \alpha_{i j}=-\frac{\pi^2 k_{b}^2 T}{3 e}\frac{d\sigma_{i j}}{d\mu} . \eeq
 These are called the Mott's relations and are very useful for discussing the physical significance of these quantities. 
 \section{Numerical plots}
 Now we use our calculated Berry curvature to comment on the nature of the variation of thermoelectric quantities with the amplitude of the light, temperature as well as with the fixed parameters of the topological insulator.

 The dependence of Nernst conductivity on the chemical potential can be obtained by controlling the band gap. If one seeks for the enhanced thermoelectric response for the topological insulators, it is important to concentrate near the Dirac point. In FIG 4, we have made numerical plots of $\alpha_{xy}$ for a fixed temperature(T=300) and a fixed optical field $e v A_0$=$0.5$eV. The plot in FIG 4 , shows significant changes with and without spin momentum unlocking terms. FIG 4a, shows the change of $\alpha_{xy}$ with chemical potential $\mu$ when the k value is lesser than the band inversion line. The fig then indicates the occurrence of only one peak since the physical response functions that we are discussing here are integrals over the Berry curvature, which in turn is a strongly peaked function with the peak around $kx=ky=0.$  Importantly, in FIG 2, we have analysed that when the k values are greater than the band crossing point, a second peak is generated(at those points where the berry curvature shows the second maximum, see FIG 1b.). Strictly speaking the second peak occurs at $\mu$$\simeq$$\pm$0.36 eV followed by a first peak at
$\mu$$\simeq$$\pm$0.06 eV. This double peak structure is new in the study of $Bi_{2}Te_{3}.$ But in \cite{TCI}, the authors have also analysed the occurrence of double peak in Nernst conductivity in case of topological crystalline insulators, which shows the uniqueness of our results. The variation of the Nernst conductivity with optical field and temperature is discussed in FIG 5. In FIG 5(a) we have shown similar plots for different optical fields. Here the red, blue and green colours of the plots indicate $e v A_{0}$=$0.3$, $0.5$, $0.9$ eV respectively. In  FIG 5(b) the Nernst conductivity is plotted with increasing temperature. The shift in the peaks with increasing temperature and optical field strength is also notable. We observe shifts of the peaks towards the Dirac point for the first peak. FIG 5b, indicates that if we keep on increasing the temperature there exists a particular temperature for which we don't achieve the two peak pattern at all. Rather one has a single broadened peak. 
From FIG 4 and FIG 5, it is clear that $\alpha_{xy}$ is always an 
odd function of chemical potential $\mu$. This nature does not diminish with
increasing of temperature, light intensity or with including higher order 
spin-momentum unlocking term.  
  \begin{figure}
  	{\centering \resizebox*{7cm}{5cm}{\includegraphics{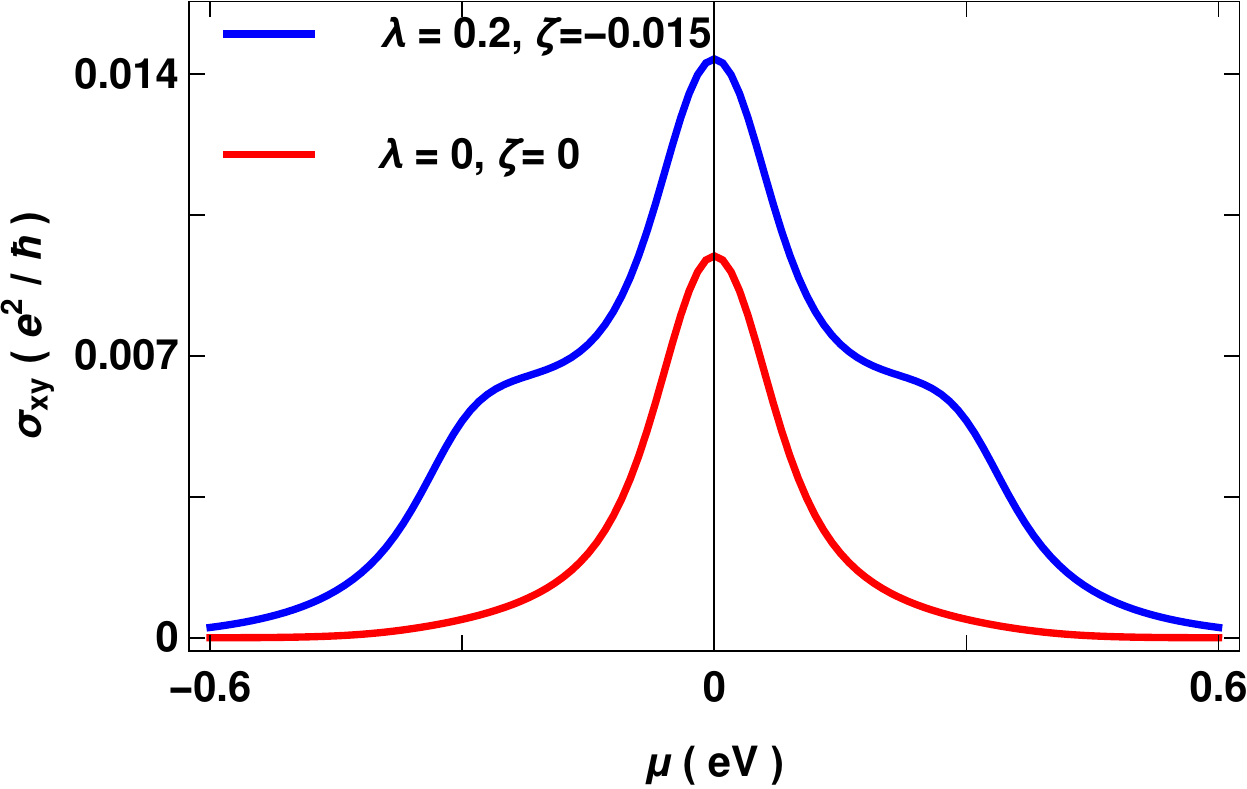}}\par}
  	\caption{(Color online). Electrical conductivity $\sigma_{xy}$ in unit of ($ e^2/\hbar$) is plotted with $\mu(eV )$  in absence and presence of higher order
  		spin-orbit coupling term. Here the optical field $e v A_0$ is fixed at 0.5 eV
  		and the temperature $T$=300K.}
  	\label{f4}
  \end{figure}
  
  \begin{figure}
  	{\centering \resizebox*{8cm}{9cm}{\includegraphics{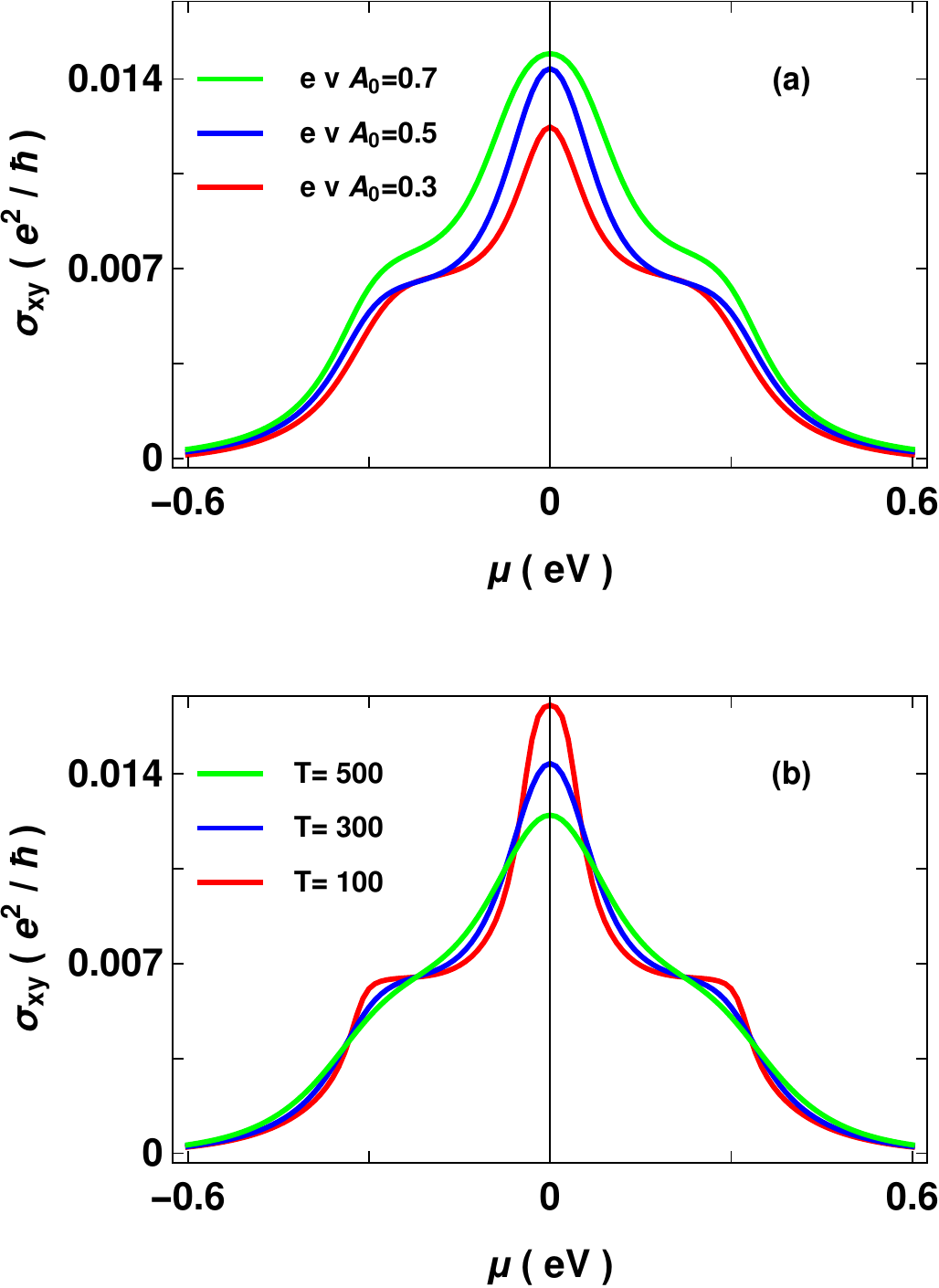}}\par}
  	\caption{(Color online). Electrical conductivity $\sigma_{xy}$ in unit of $(e^2/\hbar)$ vs $\mu(eV)$  in presence of higher order warping term$\lambda$=0.2 eV $nm^3$ and $\zeta$=-0.015 eV $nm^5$. In fig.(a), the plot is
  		for different temperature and for a fixed optical field $e v A_0$=0.5. In fig.(b), the plot is for different optical field with a fixed temperature $T$=300K.  }
  	\label{f5}
  \end{figure}
  \begin{figure}
  	{\centering \resizebox*{7cm}{5cm}{\includegraphics{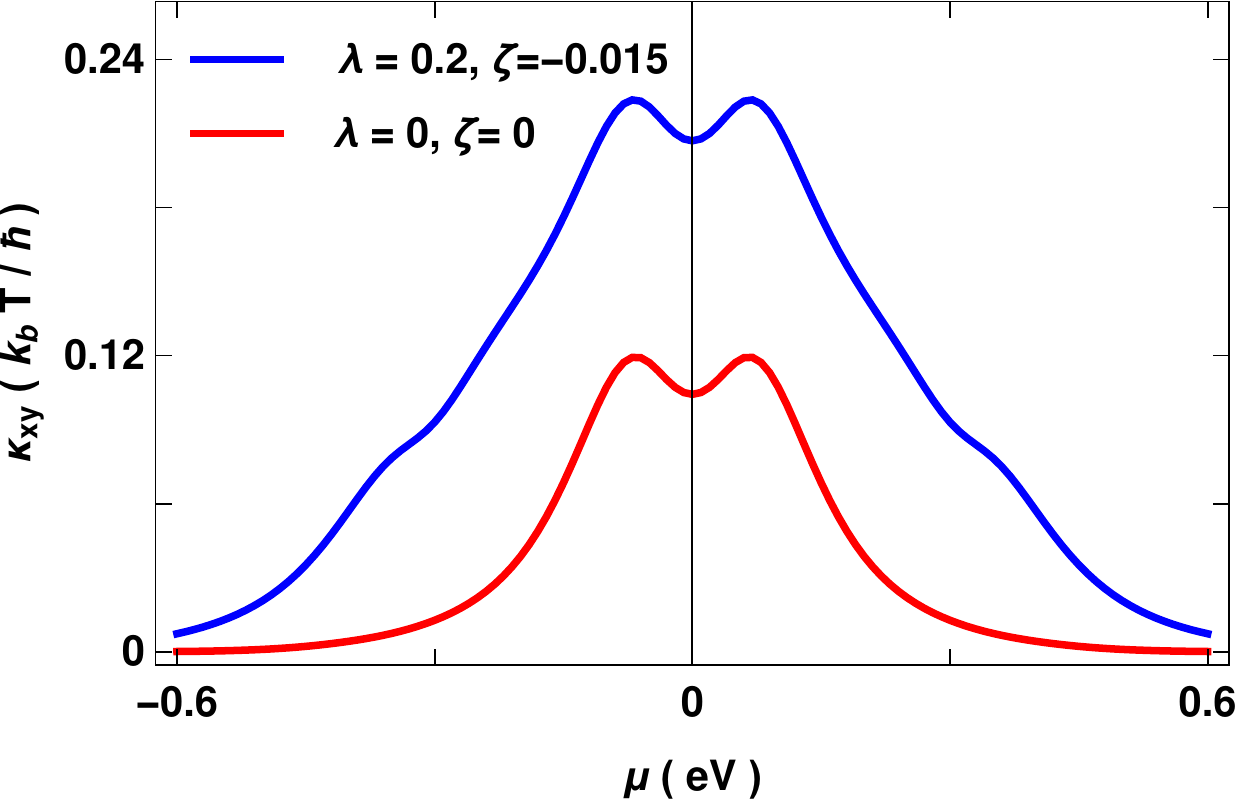}}\par}
  	\caption{(Color online).  Thermal conductivity $\kappa_{xy}$ in unit of ($ k_{b}^2 T/\hbar$) is plotted with $\mu(eV )$  in absence and presence of higher order
  		spin-orbit coupling term. Here the optical field $e v A_0$ is fixed at 0.5 eV
  		and the temperature $T$=300K.   }
  	\label{f6}
  \end{figure}
 The story also demands the similar analysis of the electrical and thermal conductivities $\sigma_{xy}$ and $\kappa_{xy}$ respectively. The numerical plot of the electrical conductivity with respect to $\mu$ is shown in FIG 6, where the red and blue line show the difference in the magnitude of conductivity when the higher order warping terms are present in the scenario. Also very minute observation shows that, the blue line in FIG 6, contains a kink around the similar value of $\mu$ as before. This kink is solely due to the presence of the spin momentum unlocking term. FIG 7(a), shows the variation of $\sigma_{xy}$ with $\mu$ for three different optical fields.  In FIG 7(b), we show the variation of the electrical conductivity with chemical potential for three different temperatures. Importantly, unlike for Nernst conductivity, the variation with temperature is opposite. Here we get higher values of the electrical conductivities for lower temperature. The maximum value of $\sigma_{xy}$ is always at $\mu$=0. From FIG 6 and FIG 7 , it is clear that $\sigma_{xy}$ is always an
even function of chemical potential $\mu$. This is the reason why we get the opposite nature of the electrical conductivity than that of the Nernst conductivity with temperature.

Finally in FIG 8, we have similar plots for $\kappa_{xy},$ which shows a larger value when higher order warping terms are switched on. In FIG 9, the plot of $\kappa_{xy}$ is analysed for three different optical fields and temperatures.

\begin{figure}
	{\centering \resizebox*{8cm}{9cm}{\includegraphics{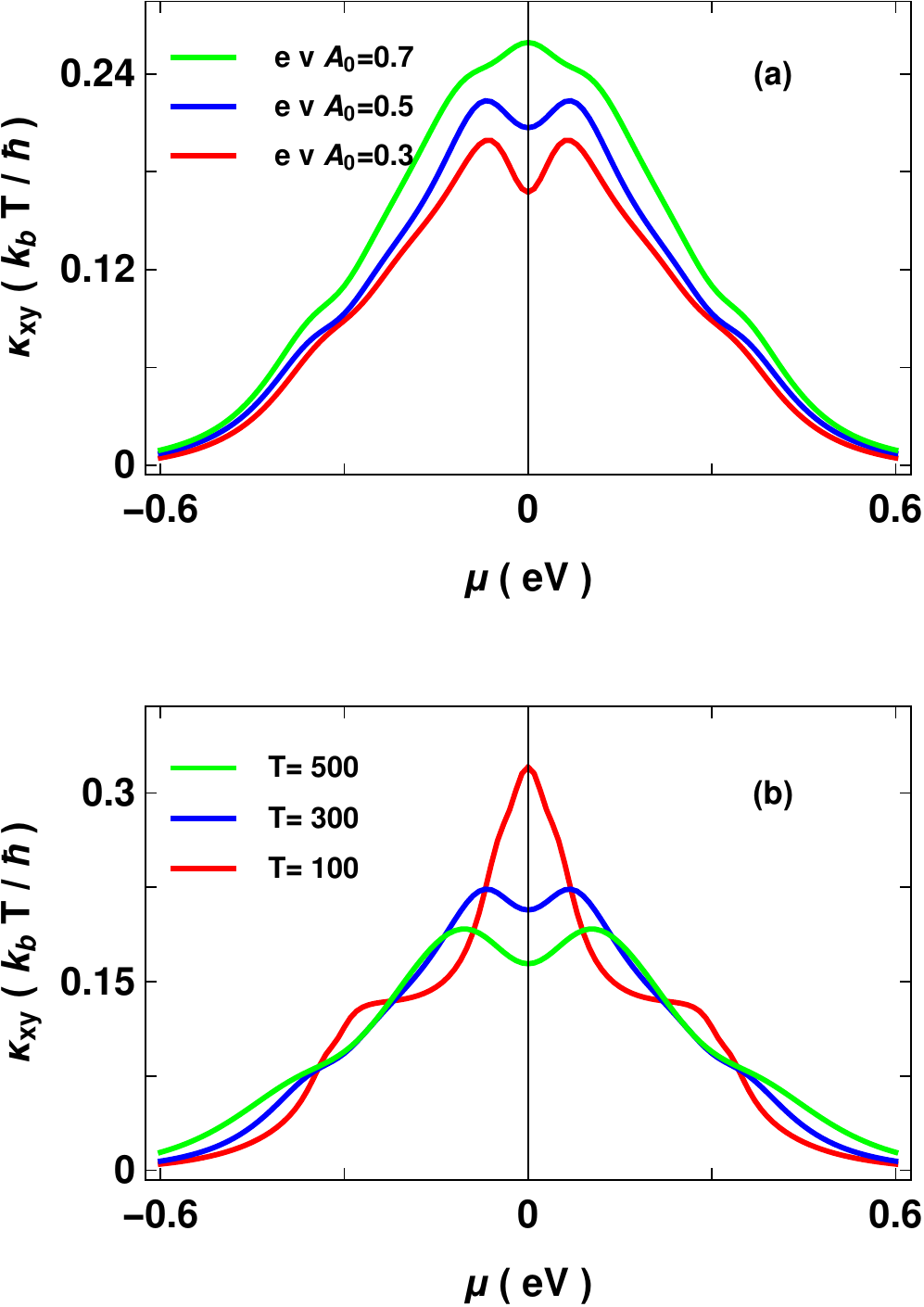}}\par}
	\caption{(Color online).Thermal conductivity $\kappa_{xy}$ in unit of ($ k_{b}^2 T/\hbar$)  vs $\mu(eV)$  in presence of higher order warping term $\lambda$=0.2 eV $nm^3$ and $\zeta$=-0.015 eV $nm^5$. In fig.(a), the plot is
		for different temperature and for a fixed optical field $e v A_0$=0.5. In fig.(b), the plot is for different optical field with a fixed temperature $T$=300K.  }
	\label{f7}
\end{figure}

 \section{Conclusion}
  In this paper our analysis is twofold. Firstly, we have used the off resonant light to the 3D topological insulator with the spin-momentum non-orthogonality.  Using Floquet theory we achieve a momentum dependent gap in the band structure. This k dependence of the optical band gap resembles the topological transition, which is a crucial point of our paper. This effectively carried forward in the Berry curvature term, which produces two maxima when varied with momentum.  To maintain the off resonant condition one needs to have a bound on the values of the momentum.   Secondly, we have analysed the  anamolous thermoelectric transport coefficients of this Floquet topological insulator in terms of the Berry curvature of the system. The Nernst and electrical conductivities are shown to vary with different temperature, different optical fields. Interestingly, in these thermoelectric coefficients, we achieve a second peak, apart from the first usual peak, due to the two maxima of the Berry curvature, which is a consequence of the spin momentum unlocking terms present in the system. Lastly, in \cite{sci}, the authors have discussed the floquet bloch states in case of 3D TI experimentally. We would like to comment that we have shown the variation of the Nernst and electrical conductivity for a wide range of temperature, which are experimentally reachable. In \cite{takehito}, the thermal and Peltier/Nernst conductivities have been discussed for magnetic topological insulators and the experiments in Refs. \cite{36,37,38} have shown good agreement with the Mott relations for temperatures up to 100 K. We hope our results will initiate some experimental verification in near future. 
 \section*{Acknowledgement}
 The authors acknowledge fruitful discussions with Dr. M. Tahir. M.S is financially supported by UGC, India. D.C  would like to acknowledge the financial support from PBC fellowship,Israel. 
 
 \appendix
 \section{Effective Hamiltonian and Floquet theory}
This section contains a general discussion on the Floquet theory and specially the computation of the  effective time independent  Hamiltonian within the off resonant approximation . With a time periodic pertubration, we have the total Hamiltonian $\mathcal{H}(t)=\mathcal{H}_{0}+V(t)$, where $\mathcal{H}_{0}$ is the static contribution and $V(t+T)=V(t)$ is the time periodic interaction. In the Fourier space, we get the Floquet Hamiltonian matrix as follows,
\begin{eqnarray}
\begin{pmatrix}
....\\
... & V_{-1} & \mathcal{H}_{-2} & V_{+1} & 0 & 0 & 0 & ...\\
... & 0 & V_{-1} & \mathcal{H}_{-1} & V_{+1} & 0 & 0 & ...\\
... & 0 & 0 & V_{-1} & \mathcal{H}_{0} & V_{+1}& 0 &...\\
... & 0 & 0 & 0 & 0 & V_{-1} & \mathcal{H}_{2} & ...\\
....
\end{pmatrix}
\end{eqnarray}
where the interaction submatrices are defined as
\begin{eqnarray}
V_{n}=\frac{1}{T}\int^{T}_{0} dt V(t) e^{-in\hbar \omega t}
\end{eqnarray}
and we have set $\mathcal{H}_{N}=\mathcal{H}_{0}+n\hbar\omega$. The corresponding  eigenstate for a given number of Fourier modes $n$, 
\begin{widetext}
\begin{eqnarray}
\phi=\begin{pmatrix}
\phi_{-n}\\
\phi_{-n+1}\\
.\\
.\\
.\\
\phi_{-1}\\
\phi_{0}\\
\phi_{1}\\
.\\
.\\
.\\
\phi_{n-1}\\
\phi_{n}
\end{pmatrix}
\end{eqnarray}
\end{widetext}
with each $\phi_{n}$ being a vector of dimensionality determined by $\mathcal{H}_{0}$. If we now restrict ourself upto order $n=1,$ we need to solve the following coupled differential eqns.
\begin{eqnarray}
\mathcal{H}_{-1}\phi_{-1}+V_{+1}\phi_{0}&=&E\phi_{-1}\\
V_{-1}\phi_{-1}+\mathcal{H}_{0}\phi_{0}+V_{+1}\phi_{+1}&=&E \phi_{0}\\
\mathcal{H}_{+1}\phi_{+1}+V_{-1}\phi_{0}&=&E\phi_{+1}
\end{eqnarray}
From the above eqns. we get,
\begin{eqnarray}
\phi_{-1}&=&(E-\mathcal{H}_{-1})^{-1}V_{+1}\phi_{0}\\
\phi_{+1}&=&(E-\mathcal{H}_{+1})^{-1}V_{-1}\phi_{0}
\end{eqnarray}
For frequencies much larger than the energy scales of the static problem, we can write,
\begin{eqnarray}
(\mathcal{H}_{0}+\frac{V_{-1}V_{+1}}{\hbar\omega}-\frac{V_{+1}V_{-1}}{\hbar \omega})\phi_{0}\simeq E \phi_{0}
\end{eqnarray}
thus the effective Hamiltonian is,
\begin{eqnarray}
\mathcal{H}_{F}\simeq \mathcal{H}_{0}+\frac{[V_{-1},V_{+1}]}{\hbar \omega}
\end{eqnarray}

\end{document}